\documentclass[sigconf]{acmart}
\AtBeginDocument{%
  \providecommand\BibTeX{{%
    Bib\TeX}}}

\setcopyright{acmlicensed}
\copyrightyear{2025}
\acmYear{2025}
\acmDOI{XXXXXXX.XXXXXXX}
\acmConference[DAC '26]{61st ACM/IEEE Design Automation Conference}{July 26-29, 2026}{Long Beach, CA, USA}
\acmISBN{978-1-4503-XXXX-X/2018/06}

\usepackage{amsmath,amssymb,amsfonts}
\usepackage{algorithmic}
\usepackage{graphicx}
\usepackage{textcomp}
\usepackage{xcolor}
\usepackage{xspace}
\usepackage{amssymb}  
\usepackage{url}
\usepackage{subcaption}
\usepackage{multirow}
\usepackage{booktabs}
\usepackage{graphicx}
\usepackage{algorithm}
\usepackage{threeparttable}
\usepackage{float}
\usepackage{pifont}
\usepackage{makecell}
\usepackage{enumitem}
\usepackage{wrapfig}
\usepackage{threeparttable}
\usepackage{flushend} 

\def\BibTeX{{\rm B\kern-.05em{\sc i\kern-.025em b}\kern-.08em
    T\kern-.1667em\lower.7ex\hbox{E}\kern-.125emX}}

\newcommand{\Name}{\texttt{TESLA}\xspace}
\begin{document}

\title{Capacitive Touchscreens at Risk: Recovering Handwritten Trajectory on Smartphone via Electromagnetic Emanations}

\author{Yukun Cheng}
\affiliation{%
  \institution{Wuhan University}
  \city{Wuhan}
  \state{Hubei}
  \country{China}
}
\email{kuin33@whu.edu.cn}

\author{Shiyu Zhu}
\affiliation{%
  \institution{Wuhan University}
  \city{Wuhan}
  \state{Hubei}
  \country{China}}
\email{zhushiyu@whu.edu.cn}

\author{Changhai Ou}
\affiliation{%
  \institution{Wuhan University}
  \city{Wuhan}
  \state{Hubei}
  \country{China}}
\email{ouchanghai@whu.edu.cn}

\author{Xingshuo Han}
\affiliation{%
  \institution{Nanjing University of Aeronautics and Astronautics}
  \city{Nanjing}
  \state{Jiangsu}
  \country{China}}
\email{xingshuo.han@nuaa.edu.cn}

\author{Yuan Li}
\affiliation{%
  \institution{National University of Defense Technology}
  \city{Changsha}
  \state{Hunan}
  \country{China}
}
\email{liyuan.margret@hotmail.com}

\author{Shihui Zheng}
\affiliation{%
  \institution{Beijing University of Posts and Telecommunications}
  \city{Beijing}
  \country{China}}
\email{shihuizh@bupt.edu.cn}

\renewcommand{\shortauthors}{Cheng et al.}

\begin{abstract}
This paper reveals and exploits a critical security vulnerability: the electromagnetic (EM) side channel of capacitive touchscreens leaks sufficient information to recover fine-grained, continuous handwriting trajectories. We present \textbf{T}ouchscreen \textbf{E}lectromagnetic \textbf{S}ide-channel \textbf{L}eakage \textbf{A}ttack (\Name), a non-contact attack framework that captures EM signals generated during on-screen writing and regresses them into two-dimensional (2D) handwriting trajectories in real time. Extensive evaluations across a variety of commercial off-the-shelf (COTS) smartphones show that \Name achieves 77\% character recognition accuracy and a Jaccard index of 0.74, demonstrating its capability to recover highly recognizable motion trajectories that closely resemble the original handwriting under realistic attack conditions.
\end{abstract}





\maketitle

\section{Introduction}
The capacitive screens of modern smartphones and tablets have evolved from basic tap-and-swipe interfaces to sophisticated platforms for rich user input. In today’s digital environment, users are increasingly relying on handwriting for sensitive tasks, such as recording private notes, composing multilingual messages, and most critically, providing handwritten signatures for identity verification and transaction authorization in financial and legal applications. Although this interaction mode enhances usability and personalization, it inadvertently introduces serious privacy vulnerabilities. This raises a fundamental security question: can the highly sensitive trajectories of fingertip movements on the capacitive screen be covertly recovered through physical side channels?

Previous research on side-channel attacks targeting handwriting has investigated several attack vectors. Sensor-based approaches have leveraged motion sensors, including magnetometers \cite{FarrukhYXYWC21} and gyroscopes \cite{TaktakTK17}, to infer writing patterns by analyzing subtle device vibrations and orientation changes during the writing process. Meanwhile, other works have examined the feasibility of acoustic side channels \cite{ZhangYTSTX15, YuJN16, YuJN20}, utilizing the device's microphone to capture writing-related sounds and reconstruct the written content.

However, the practical effectiveness of these existing approaches is often subject to several limitations. (i) \textit{Access authorization requirements}. Sensor-based methods \cite{FarrukhYXYWC21, TaktakTK17} typically require user authorization to install malicious software on the smartphone or stylus to access sensor data. (ii) \textit{Laboratory-condition dependency}. Several acoustic attacks assume that the victim writes in specific print styles, which significantly undermines the realism and generalizability of the attack model \cite{YuJN16, YuJN20}. (iii) \textit{Specialized hardware dependencies.} Reliance on custom-designed measurement hardware increases assumptions regarding the attacker's capabilities and complicates attack deployment \cite{ZhangWG0Z25, WeiZ15}.

\begin{table*}[!tbp]
\caption{Comparison with representative side-channel attacks targeting handwriting.}
\resizebox{\linewidth}{!}{\begin{tabular}{c|l|l|c}
\Xhline{1.5pt}
\textbf{Attack Method}                              & \multicolumn{1}{c|}{\textbf{Side Channel and Information}}                                                                & \multicolumn{1}{c|}{\textbf{Measurement Approach}}                                                                               &  \textbf{Specific Laboratory-condition}                    \\ \Xhline{1.5pt}
$S^3$ \protect\cite{FarrukhYXYWC21}             & Magnetic data during the Apple pencil writing   & Magnetometer of victim’s iPad   & Prior knowledge of the victim’s writing behavior     \\ \hline
WritingHacker \protect\cite{YuJN20}  & Sound of the victim’s handwriting & Recorder of another nearby smartphone  & Victim’s handwriting follows certain print style \\ \hline
RadSee \protect\cite{ZhangWG0Z25} & Radio signals during the victim’s handwriting & Specialized radar hardware & None           \\ \hline
mtrack \protect\cite{WeiZ15}      & mmWave during the victim’s handwriting  & Software-defined radio platform & None      \\ \hline
\textbf{\Name}                                      & EM emanations during the victim’s handwriting & COTS EM probe & None \\ \Xhline{1.5pt}
\end{tabular}}
\label{tab: compare}
\end{table*}

In this paper, we present \textbf{T}ouchscreen \textbf{E}lectromagnetic \textbf{S}ide-channel \textbf{L}eakage \textbf{A}ttack (\Name), a practical and novel contactless attack that exploits a previously overlooked physical vector, the electromagnetic (EM) side channel emitted by the capacitive touchscreen during user interaction, to reconstruct handwriting trajectories. An adversary can simply place a probe in proximity (e.g., within 15 cm) to a victim's smartphone to capture EM emanations generated during touch events, thereby recovering sensitive handwritten input trajectory. This leakage originates from the interplay between the Human Coupling Effect and the sequential Scan Driving Method (SDM), which imparts location-dependent characteristics to the EM signals. \Name captures the continuous evolution of these signals and employs a transformer-based model to directly map the collected EM data to the two-dimensional (2D) trajectory of the user's handwriting.

Compared to existing methods, \Name offers the following advantages, as detailed in Table~\ref{tab: compare}. (i) Unlike previous works that rely on sensor data from the target device, e.g., magnetometer readings \cite{FarrukhYXYWC21} or gyroscope measurements \cite{TaktakTK17}, \Name executes the attack by simply placing a measurement probe close to the target device, without device access authorization. (ii) \Name demonstrates greater robustness with respect to variations in users' writing habits compared to existing attacks \cite{YuJN16, YuJN20}, which require victims to write in specific printed styles. (iii) \Name does not require custom-designed hardware, such as radar antennas operating at specific frequencies \cite{ZhangWG0Z25, WeiZ15}, to measure leakage signals. Instead, attackers can capture EM emanations using commercial off-the-shelf (COTS) equipment costing less than \$100.

The main contributions of this paper are as follows:

\begin{itemize}
\item \textbf{We demonstrate a novel class of EM side-channel vulnerability,} 
showing that EM emanations from capacitive touchscreens can leak the continuous fine-grained trajectory of finger movements.

\item \textbf{We design and implement \Name,} 
the first attack framework that exploits this vulnerability. By leveraging a transformer network model, we translate raw EM signal streams into accurate 2D motion paths, enabling the reconstruction of handwritten trajectories.  

\item \textbf{We conduct a comprehensive evaluation across multiple COTS smartphones,} 
(iPhone X, Xiaomi 10 Pro, Samsung S10, and Huawei Mate 30 Pro), demonstrating the practical effectiveness of the attack in real-world scenarios and its robustness against various practical impact factors.
\end{itemize}

\section{Preliminaries}
\subsection{Capacitive Touchscreen}
Modern smartphones predominantly use capacitive touchscreens, which detect touch by measuring changes in mutual capacitance on a sensor grid composed of transmitting (TX) and receiving (RX) electrodes \cite{kwon2018capacitive}. An analog front-end (AFE) IC sends an excitation signal to the TX lines, creating an electrostatic field; a finger approaching the screen disturbs this field, causing a change in mutual capacitance that the AFE IC detects to pinpoint the touch.

To manage this sensing process, most smartphones employ the Scan Driving Method (SDM) for its simple design and high speed \cite{wang2022ghosttouch}. In SDM, the excitation signal is sent sequentially to each TX electrode via a demultiplexer. This sequential scanning mechanism in SDM is the foundation for our proposed side-channel vulnerability. Because the excitation signal scans different screen positions at different times, it inherently creates predictable time delays corresponding to each location. This timing feature provides an exploitable side channel for inferring touch positions through EM emanation analysis.

\subsection{Human Coupling Effect}
\begin{figure}[t]
\centering
\subfloat[]{\includegraphics[width=.5\linewidth]{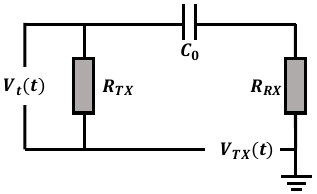}\label{no_touching}}%
\subfloat[]{\includegraphics[width=.5\linewidth]{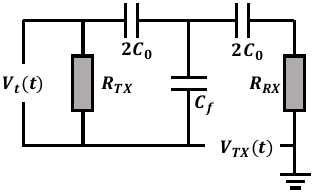}\label{touching}}
\caption{Illustration of human coupling effect in touch screen. (a) Without touching. (b) With touching.}
\label{finger_coupling}
\vspace{-10pt}
\end{figure}

Figure \ref{finger_coupling} illustrates the human-coupling effect, which significantly influences the driving voltage $V_t(t)$ of the electromagnetic emissions of a capacitive touchscreen  during touch events \cite{NiZZ23, jin2021periscope}.

When no touch is present, as shown in Figure \ref{no_touching}, the touch screen can be modeled by a capacitance $C_0$ between the TX and RX electrodes. Under this condition, the driving voltage $V_t(t)$ is:
\begin{equation}
  V_t(t) = V_{TX}(t) \cdot \frac{R_{TX}}{R_{TX} + R_{RX} + \frac{1}{j 2 \pi f C_0}},
\end{equation}
where $V_{TX}(t)$ is the excitation signal, $R_{TX}$ and $R_{RX}$ are the resistors at TX and RX electrodes, $f$ is the current frequency, and $j$ is the imaginary unit.

When a finger touches the screen, as shown in Figure \ref{touching}, it introduces an additional capacitance $C_f$. This coupling alters the system's impedance, and the driving voltage $V_t(t)$ becomes:
\begin{equation}
  V_t(t) = V_{TX}(t) \cdot \frac{R_{TX}}{R_{TX} + \frac{1}{j 4 \pi f C_0} + \Delta Z_f(t)},
\end{equation}
where the impedance change $\Delta Z_f(t)$ is defined as:
\begin{equation}
  \Delta Z_f(t) = \frac{1}{\left( \frac{1}{1/{j 2 \pi f \Delta C_f}} + \frac{1}{1/{j 4 \pi f C_0 + R_{RX}} }\right)}.
\end{equation}

As the finger approaches the screen, $C_f$ increases, strengthening the coupling. This consequently reduces the impedance $\Delta Z_f(t)$, which increases the driving voltage $V_t(t)$ and amplifies the electromagnetic emissions. This distance-dependent signal variation provides a potential side channel for detecting touch interactions.

\subsection{Jaccard Index}
The Jaccard index was introduced by Swiss Jaccard to compare the distribution of flora across different geographic regions \cite{jaccard1912distribution}. It has since been widely adopted in various fields, including computer vision and data mining, for measuring the similarity between sample sets. Mathematically, the Jaccard index quantifies the similarity between two finite sample sets $A$ and $B$, and is defined as the ratio of the size of their intersection to the size of their union:

\begin{equation}
J(A, B) = \frac{|A \cap B|}{|A \cup B|} = \frac{|A \cap B|}{|A| + |B| - |A \cap B|}.
\end{equation}

The Jaccard index ranges from 0 to 1. A value of 1 indicates a perfect match, whereas a value of 0 indicates no overlap or similarity between the two sets. In the context of our attack, a higher Jaccard index reflects a more accurate recovery of the user's biometric handwriting features.

\subsection{Related Works}
Side channel analysis on smart devices primarily targets sensitive user inputs, ranging from discrete PINs and passcodes to general keyboard entries, and ultimately to complex, continuous inputs such as handwriting trajectories. These attacks have exploited various physical leakage vectors—including acoustic, radio-frequency (RF) signals, built-in sensors, power consumption, and EM emanations to compromise the input confidentiality.

\noindent\textbf{PIN Recovery.}
A substantial body of research has focused on recovering short, fixed-pattern inputs like screen unlock PINs. For instance, researchers have successfully inferred PINs by analyzing data from built-in ambient light sensors \cite{spreitzer2014pin} or smartphone accelerometers \cite{AvivSBS12}. Cronin et al. achieved high accuracy in identifying 4-digit passcodes by monitoring power traces through a compromised USB charger \cite{Cronin0YW21}. Similarly, EM emanations from touchscreens have been leveraged in attacks such as Periscope to recover 4-digit PINs \cite{jin2021periscope}. These studies collectively demonstrate that discrete and simple inputs are vulnerable across multiple side-channel modalities.

\noindent\textbf{Keystroke Inference.}
Beyond basic PINs, researchers have extended their focus to keyboard inputs to capture messages or passwords. Pioneering work utilized accelerometer and gyroscope data to infer keystrokes based on distinctive vibration patterns \cite{CaiC11}. Ni et al. further demonstrated keystroke inference using another phone’s microphone and magnetometer \cite{NiZZLYWXLZ23}, while other studies have reconstructed typed content by analyzing EM leakage associated with screen updates and graphics rendering operations \cite{Zhuoran2021Screen, Zhan2022Graphics}.

\noindent\textbf{Handwriting reconstruction.}
More sophisticated, continuous inputs such as handwriting trajectories have also been targeted using diverse attack vectors. Sensor-based approaches include $S^{3}$, which exploits an iPad’s built-in magnetometer to track the embedded magnets in a stylus (e.g., Apple Pencil) and reconstruct written content \cite{FarrukhYXYWC21}. Other methods employ gyroscope signals from smartphones to recognize 3D “air writing” gestures, where the device itself is moved as an input tool \cite{TaktakTK17}. Acoustic side channels have similarly been investigated: WritingHacker showed that a mobile device placed on the desk can record the acoustic signatures of print-style handwriting and recognize words by extracting features such as stroke count \cite{YuJN20}. Finally, advanced RF systems have been employed. mTrack uses 60 GHz millimeter-wave radios to create a virtual trackpad, enabling passive tracking of fine-grained pen movements through phase shift analysis \cite{WeiZ15}. RadSee utilizes a 6 GHz FMCW radar system to detect hand motions and recognize handwritten letters, even through obstacles such as walls \cite{ZhangWG0Z25}.

\section{Touchscreen Electromagnetic Side-channel Leakage Attack}
This section presents our proposed Touchscreen Electromagnetic Side-channel Leakage Attack (\Name). We begin by outlining the threat model, followed by an analysis of the touchscreen leakage mechanism. Finally, we provide an overview of the attack framework.

\subsection{Threat Model}
We consider two practical attack scenarios in both public (e.g., a library) and private (e.g., a meeting room) environments, these settings that are common in daily life and have been adopted in prior studies \cite{jin2021periscope, Zhuoran2021Screen}. In these scenarios, a victim places the smartphone on a table to perform sensitive handwritten input, such as signing a legal document, authorizing a financial transaction, or writing a private note. The attacker has previously deployed a concealed EM probe in close proximity (e.g., beneath the tabletop or inside a nearby bag), which non-intrusively captures the EM emanations produced by the victim's continuous finger movements on the touchscreen.

Regarding the attacker's capabilities, we assume that they can capture the EM signals emitted by the victim's device during the writing process. Additionally, the attacker can identify the victim's phone model through network traffic analysis or Bluetooth sniffing, which is a commonly accepted assumption in related work \cite{jin2021periscope, WangMY0SX22}. However, the attacker does not require physical access to the device and cannot install malicious software. They also cannot visually observe the screen or hand movements. Furthermore, the attack is passive and does not rely on restrictive conditions, such as requiring the victim to write in a certain printed style.

\begin{figure}[!tbp]
  \centering
  \includegraphics[width=\linewidth]{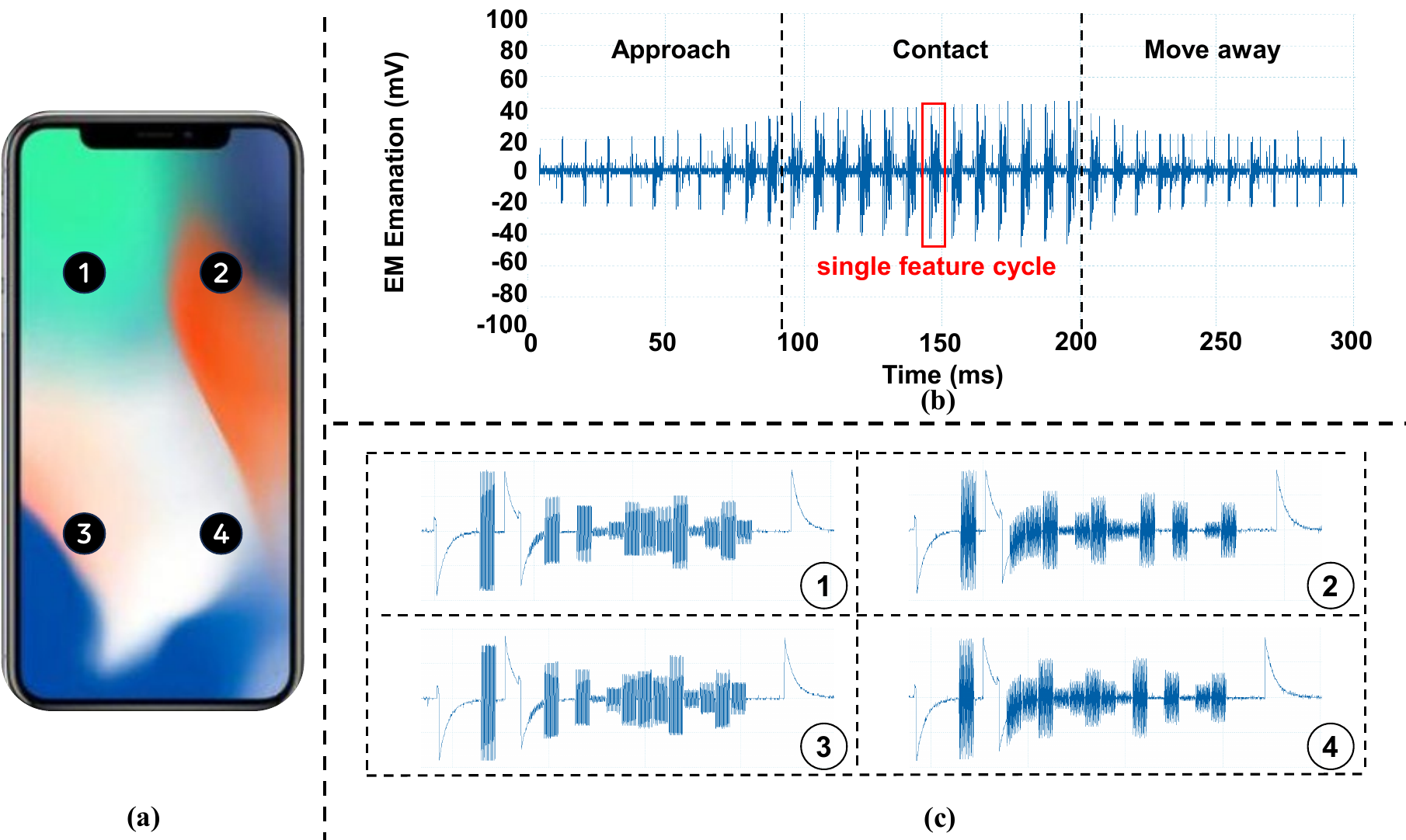}
  \caption{EM emanation measurements for touch interactions on iPhone X. (a) Four different touch positions. (b) EM emanation measurements for single touch interaction. (c) Details of EM emanation measurements within a single feature cycle for different touch positions.}\label{fig: waveform}
  \vspace{-10pt}
\end{figure}

\subsection{Leakage Mechanism Analysis}
\label{sec: leakage}
To understand the physical origin of the leakage, we conducted an empirical study by placing an EM probe in close proximity to a target smartphone (e.g., an iPhone X) and recording the EM signals using an oscilloscope.

Our first observation is that the EM signal intensity correlates directly with the user's finger interaction. As shown in Figure \ref {fig: waveform}(b), the signal amplitude gradually increases as the finger approaches the screen, peaks during contact, and decreases as the finger moves away. When further analyzing the signal within a single feature cycle, as shown in Figure \ref{fig: waveform}(c), we observe that the waveform exhibits distinct, location-dependent characteristics. Touches at different horizontal positions produce pronounced variations in the time-domain features of the signal, whereas touches at different vertical positions result in more subtle differences. This demonstrates that the EM signal is spatially dependent and, critically for our work, indicates that continuous movements (e.g., handwriting) generate a continuously evolving, location-specific signal waveform.

Further analysis confirms the hardware-level origin of this leakage. As shown in Table \ref{table: frequency}, we measured the frequency of the signal’s cyclical features and found it to be identical to the device’s touch sampling frequency (e.g., 120Hz on an iPhone X) and independent of the screen refresh frequency (e.g., 60Hz), as documented in official technical specifications \cite{APPLE23, MI25, Sam21, Hua22}. This finding rules out software-level visual rendering, which is associated with screen refresh frequency and has been identified as the source of leakage in classical side-channel attacks on mobile phones \cite{Cronin0YW21}, as the cause. Instead, the leakage arises as a direct byproduct of the touchscreen’s underlying SDM. In SDM, the TX electrodes are activated sequentially, introducing location-dependent time delays into the electrical scanning process. When the human coupling effect interacts with this sequential scan, it generates a unique, time-varying EM signature that accurately reflects the finger’s physical coordinates, thereby leaking a continuous stream of positional data.

\begin{table}[t]
\centering
\caption{The screen-related parameters of the smartphones and the frequency of the cyclical feature.}
\renewcommand{\arraystretch}{1}
\resizebox{\linewidth}{!}{
\begin{tabular}{c|c|c|c}
\Xhline{1.5pt}
\textbf{Smartphone} & \textbf{\begin{tabular}[c]{@{}c@{}}Screen Refresh \\ Frequency (Hz)\end{tabular}} & \textbf{\begin{tabular}[c]{@{}c@{}}Touch Sampling \\ Frequency (Hz)\end{tabular}} & \textbf{\begin{tabular}[c]{@{}c@{}}Cyclical Feature \\ Frequency (Hz)\end{tabular}} \\ \Xhline{1.5pt}
iPhone 7            & 60                                                                                & 60                                                                                & 60                                                                                  \\
iPhone X            & 60                                                                                & 120                                                                                & 120                                                                                  \\
Xiaomi 10 Pro       & 90                                                                                & 180                                                                               & 180                                                                                 \\
Samsung S10         & 60                                                                                & 120                                                                               & 120                                                                                 \\
Huawei Mate30 Pro   & 60                                                                                & 120                                                                               & 120                                                                                 \\ \Xhline{1.5pt}

\end{tabular}}
\label{table: frequency}
\end{table}

\subsection{\Name Attack Framework}

\begin{figure*}[!tbp]
  \centering
  \includegraphics[width=\linewidth]{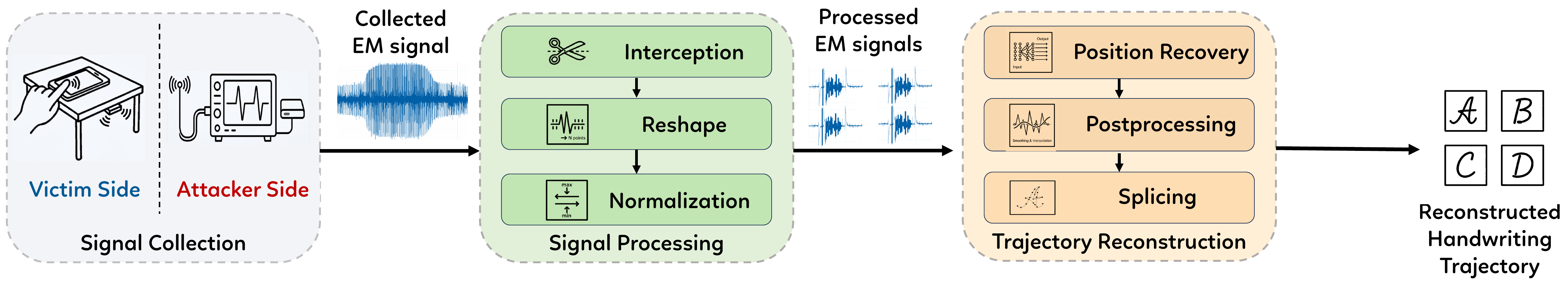}
  \caption{Overview of the \Name.}\label{fig: Overview}
\end{figure*}

As illustrated in Figure \ref{fig: Overview}, the \Name attack is structured as a three-stage pipeline: signal collection, signal preprocessing, and trajectory reconstruction.

The signal collection stage originates on the victim side, where a covertly deployed EM probe captures real-time analog EM leakage signals generated during the user's handwriting interactions. These raw signals are continuously transmitted to the attacker’s side via cable, Wi-Fi, or Bluetooth modules, where they are digitized, recorded, and preserved for subsequent processing and trajectory recovery.

In the signal preprocessing stage, the raw and noisy EM signal stream is transformed into standardized feature vectors suitable for deep learning models. This process begins with an interception step, in which the continuous signal is segmented into discrete windows, each corresponding to a single feature cycle based on the device's known touch sampling frequency. Subsequently, a reshaping operation is applied, employing up-sampling or down-sampling techniques to ensure uniform length across all segments. Finally, Z-score normalization is performed on the reshaped segments to standardize their amplitude and statistical distribution, thereby enhancing input stability for the deep neural network.

The final trajectory reconstruction stage consists of a three-step process designed to transform processed signals into a clean and coherent path. First, in the position recovery step, the processed and normalized EM signals are sequentially fed into a transformer-based deep neural network. This network is trained to estimate a sequence of touch positions, producing 2D coordinates for each time step. Next, the raw coordinate sequence undergoes post-processing, during which smoothing techniques, such as filtering and interpolation, are applied to reduce noise and mitigate jitter introduced during the initial recovery phase. Finally, in the splicing step, the refined positional estimates are connected in their correct temporal order to generate the final reconstructed handwritten trajectory.

\section{Experimental Evaluation}
\label{sec: setup}
\subsection{Experimental Setup}
\noindent\textbf{Target devices and environments.} 
To ensure the broad applicability of our findings, we evaluated \Name on four COTS smartphones (i.e., iPhone X, Xiaomi 10 Pro, Samsung S10, and Huawei Mate 30Pro) across two representative environments: a private meeting room environment and a public library environment. We began our analysis with the iPhone X in a private setting (i.e., the probe was concealed beneath a 5 cm thick table) to establish the effectiveness of \Name, followed by presenting results from the remaining devices and the public environment.

\noindent\textbf{Data collection.} 
Our data were collected from 10 participants who performed handwriting tasks, including writing individual characters and complete words, in similar private and public environments, with the smartphone naturally placed on the table. A Fosttek NFP-One P1 EM measurement probe was used to capture EM signals, while an STM32F407 microcontroller unit (MCU) served as the data sampling and storage device. Additionally, a PicoScope 5444D oscilloscope was employed as a supplementary tool for real-time signal observation.

\noindent\textbf{Data processing.}
The collected data were processed on a high-performance Dell Precision server equipped with an Intel Xeon Silver 4210R CPU running at 2.40 GHz, 32 GB of DDR5 memory, a 1 TB Samsung SSD-980-PRO storage drive, and an NVIDIA GeForce RTX 3090 GPU. The deep learning framework was implemented using Keras 2.9.0 as the frontend and TensorFlow 2.9.1 as the backend, with CUDA 11.7 enabling GPU acceleration during model training. The architecture of our transformer model is detailed in Table \ref{table: model}. It begins with a 1D convolutional (Conv1D) front-end for feature extraction, followed by a Transformer Encoder that processes the resulting sequence with a model dimension ($d_{model}$) of 256, a feed-forward dimension ($d_{ff}$) of 1024, and a maximum sequence length ($max\_len$) of 5000. An attention pooling layer aggregates the encoder outputs, which are then passed through a multi-layer perceptron (MLP) composed of three fully connected layers (FC) for signal classification.

\begin{table}[t]
\centering
\caption{Transformer model architecture.}
\renewcommand{\arraystretch}{1}
\resizebox{\linewidth}{!}{
\begin{tabular}{c|c|c}
\Xhline{1.5pt}
\textbf{Layer Group}   & \textbf{Output Shape} & \textbf{Key Parameters / Structure}   \\ \Xhline{1.5pt}
Input                 &  (1, $N_{input}$)            &                     -                                                                                            \\ \hline
Convolutional Front-End & (256, 445)            & \begin{tabular}[c]{@{}c@{}}Conv1d(1, 64, kernel=7, stride=2) -\textgreater \\ Conv1d(64, 256, kernel=5, stride=2)\end{tabular} \\ \hline
Positional Encoding    & (445, 256)            & $d_{model}$=256, $max\_len$=5000                                                                             \\ \hline
Transformer Encoder    & (445, 256)            &  $d_{ff}$=1024                                                                                                        \\ \hline
Attention Pool         & (256)                 & FC(256-\textgreater{}128) -\textgreater FC(128-\textgreater{}1)\\ \hline
Classifier (MLP)       & ($N_{class}$)            & FC(256-\textgreater{}512) -\textgreater FC(512-\textgreater{}256) -\textgreater FC(256-\textgreater{}$N_{class}$)                                                                                      \\ \Xhline{1.5pt}

\end{tabular}}
\label{table: model}
\end{table}

\subsection{Touch Position Recovery}
\begin{figure}[!tbp]
  \centering
  \includegraphics[width=\linewidth]{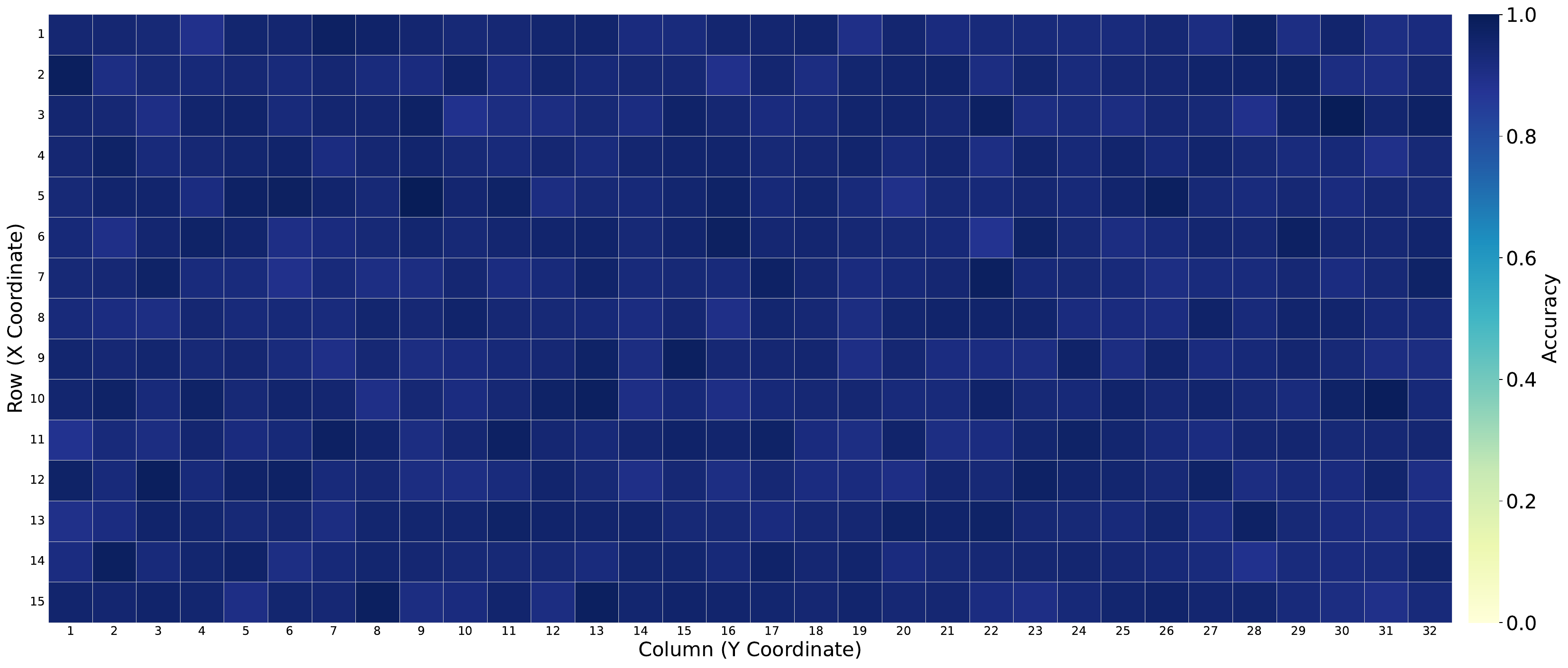}
  \caption{Evaluation results of touch position recovery.}\label{fig: accuracy}
  \vspace{0pt}
\end{figure}

We first evaluate the model's ability to recover static touch positions. The smartphone screen is divided into a $32 \times 15$ grid, resulting in 480 distinct zones. For each zone, we collected EM signals during a 5-second touch interaction and processed 1,000 samples, yielding a dataset of 480,000 samples that comprehensively capture spatial variations across the screen. The dataset was partitioned into training and testing subsets in an 80:20 ratio (i.e., 384,000 training samples and 96,000 testing samples) to ensure representative coverage of touch positions while preventing data leakage between training and testing sets. As shown in Figure \ref{fig: accuracy}, \Name achieves outstanding performance in recovering touch positions, attaining an exceptional overall accuracy of 94.08\% in identifying the correct 32×15 grid region. This demonstrates the model’s high spatial resolution.

\subsection{Handwriting Trajectory Recovery}
We next evaluated the performance of \Name in recovering continuous handwriting trajectories. Given that users typically write individual words rather than full sentences when handwriting on smart devices, our evaluation focuses primarily on character-level and word-level trajectory recovery.  

\begin{figure}[t]
\centering
\subfloat[]{\includegraphics[width=.5\linewidth]{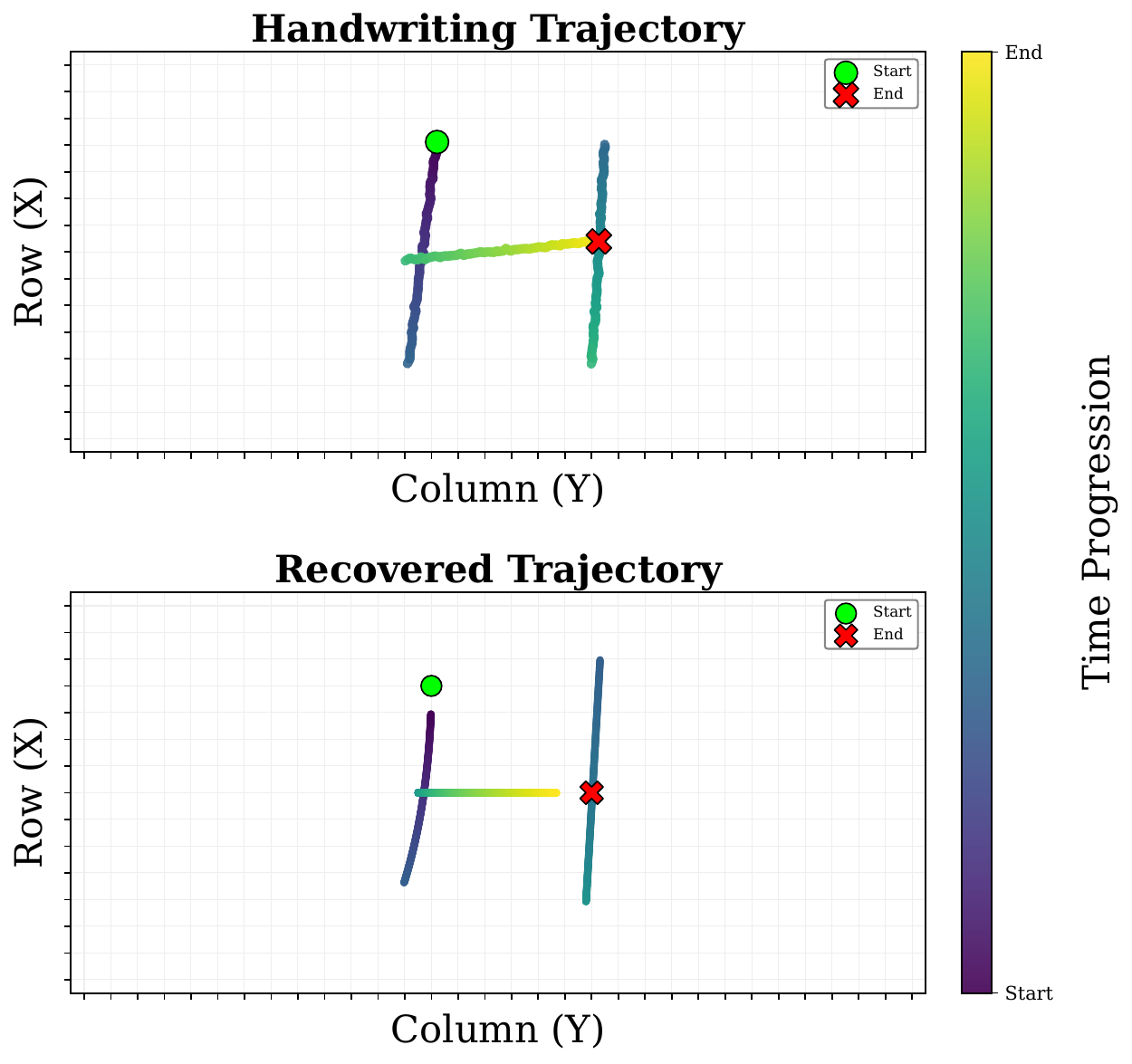}\label{fig: letter}}%
\subfloat[]{\includegraphics[width=.5\linewidth]{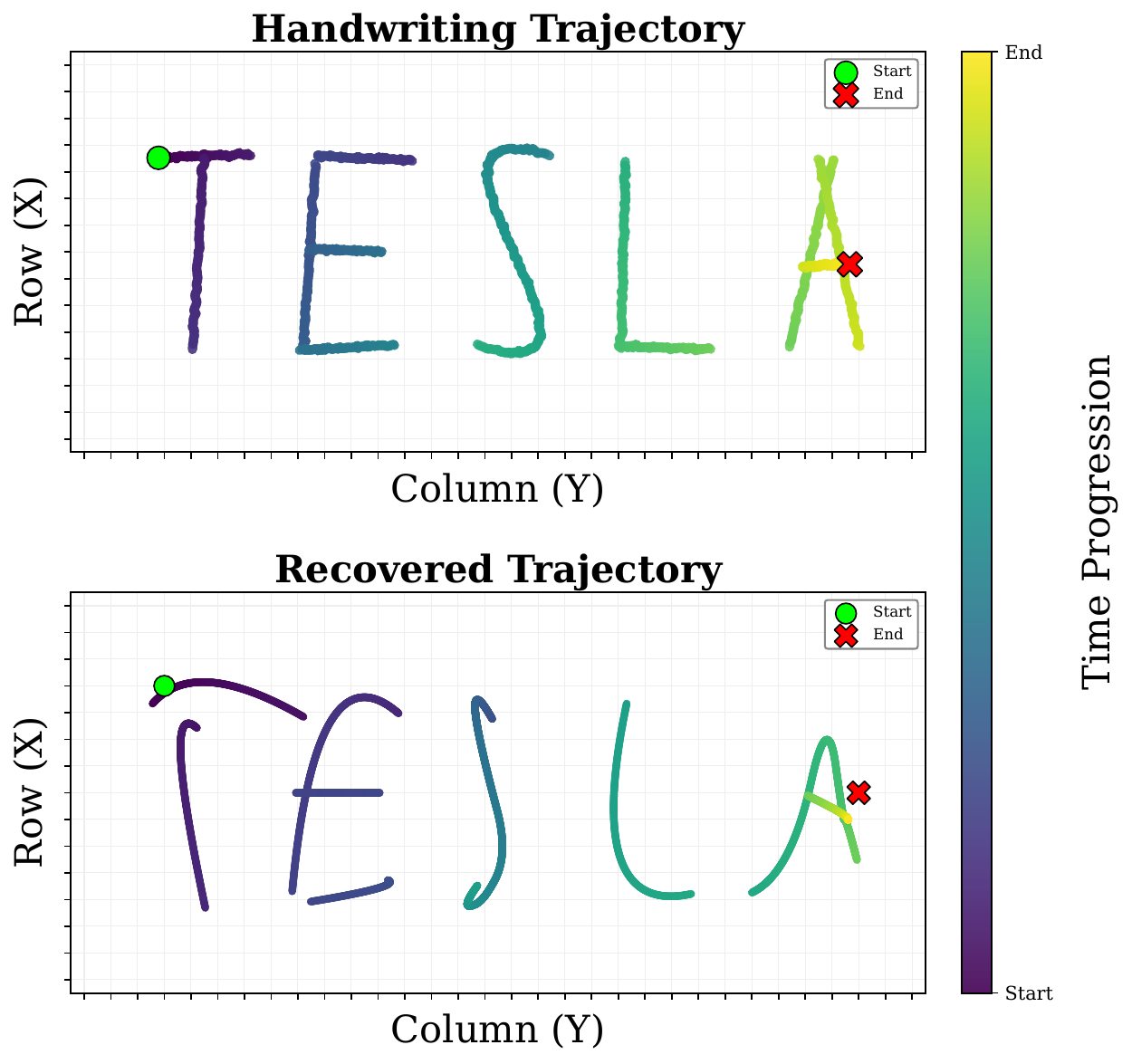}\label{fig: word}}
\caption{Comparisons of handwriting trajectory and recovered trajectory. (a) Character-level. (b) Word-level.}
\label{fig: handwriting}
\end{figure}

\begin{figure}[!tbp]
  \centering
  \includegraphics[width=\linewidth]{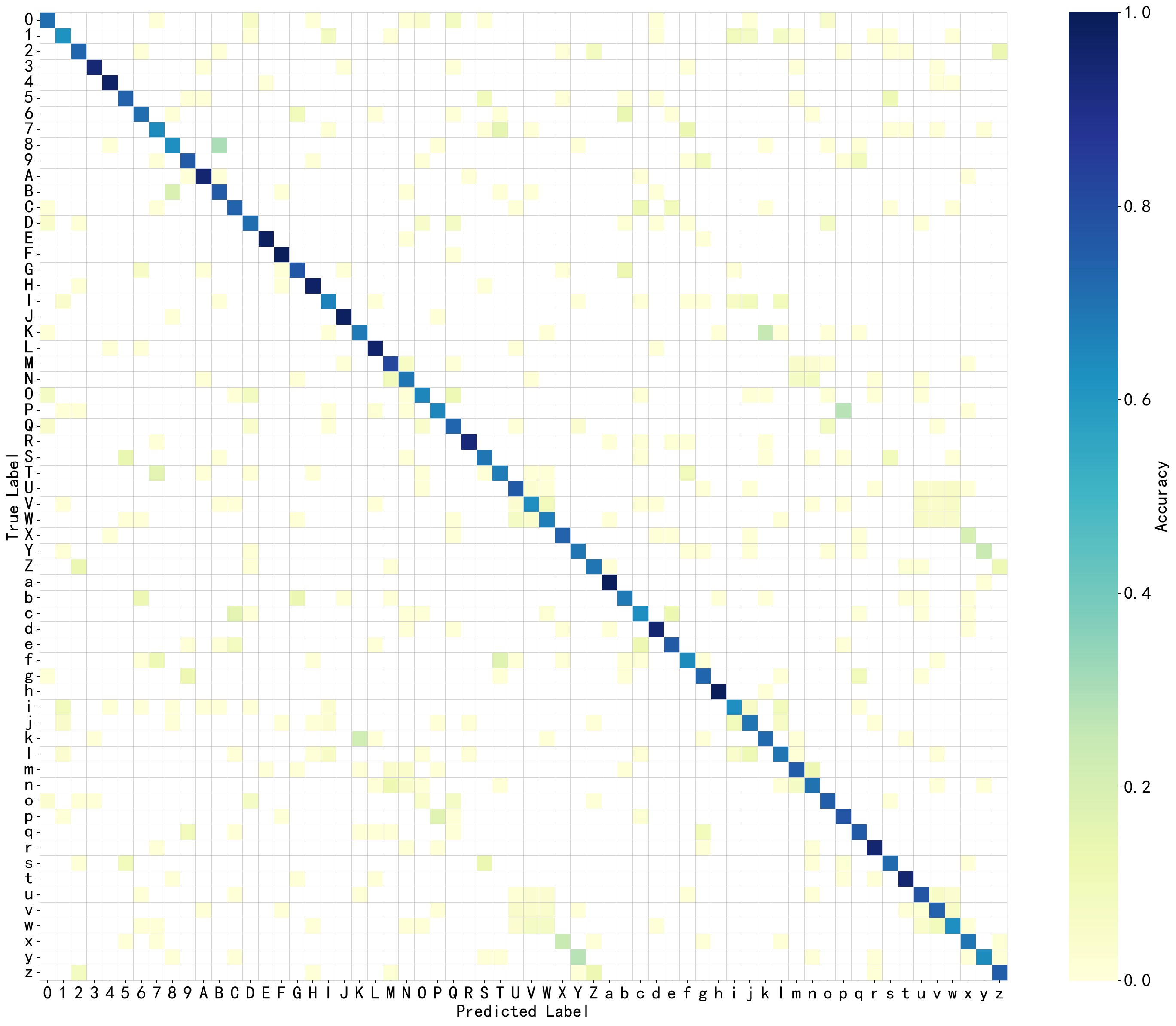}
  \caption{Confusion matrix of character recognition results.}\label{fig: confusion}
  \vspace{-10pt}
\end{figure}

\noindent\textbf{Character-level recovery.} 
We first assessed the recovery of individual characters. As shown in Figure \ref{fig: handwriting}(a), \Name achieves high-fidelity reconstruction of single-character trajectories. The recovered trajectories accurately preserve the distinct shapes of the original handwriting. We then conducted a quantitative evaluation under two distinct threat scenarios: content recovery (e.g., stealing notes) and biometric trajectory recovery (e.g., signature forgery). In the content recovery scenario, we input 6,200 recovered trajectories (i.e., 100 trajectories per character) into a third-party optical character recognition (OCR) tool Tesseract-OCR \cite{TessOverview}. As shown in Figure \ref{fig: confusion}, the resulting confusion matrix across all 62 characters (A–Z, a–z, and 0–9) yielded an average recognition accuracy of 76.77\%. Errors primarily arise from two understandable sources: non-standard writing variations (e.g., '7' vs. 'T', '9' vs. 'q') and visually similar character pairs (e.g., '0' vs. 'O', 'I' vs. 'l'). In the biometric recovery scenario, we measured geometric similarity between original and recovered trajectories using the Jaccard index. Across 100 random sample pairs, the average Jaccard index similarity reached 0.7374, indicating well spatial overlap.

\begin{table}[t]
\centering
\caption{Evaluation results of handwriting trajectory recovery.}
\renewcommand{\arraystretch}{1}
\resizebox{\linewidth}{!}{
\begin{tabular}{c|c|c}
\Xhline{1.5pt}
\textbf{Recovery Level}   & \textbf{Recognition Accuracy} & \textbf{Jaccard Index}   \\ \Xhline{1.5pt}
Character-level                &  0.7677            &                     0.7374                                                                                            \\ 
Word-level            & 0.5834 &                     0.6661 \\ 
\Xhline{1.5pt}

\end{tabular}}
\label{table: handwriting}
\end{table}

\noindent\textbf{Word-level recovery.} 
Word-level recovery results are presented in Figure \ref{fig: handwriting}(b). For word-level recovery, we performed five consecutive attempts using a third-party recognition tool and adopted the top-5 recovery success rate as the evaluation metric to evaluate the recognizability of the recovered trajectories, defined as a successful attack if any of the first five outputs correctly match the target word. As shown in Table \ref{table: handwriting}, \Name achieves 58.34\% top-5 accuracy, demonstrating that the attack remains effective for practical, real-world word-level recognition tasks. We further applied the Jaccard index to measure trajectory similarity, which remained high at 0.6661. This slight performance degradation compared to character-level recovery is likely attributable to cumulative errors in recovering multiple characters sequentially, as well as to character truncation operations introduced during postprocessing to ensure the independence of each character’s reconstructed trajectory within a word.

\subsection{Impact Factors}
To demonstrate the robustness and practicality of \Name, we evaluated its performance under several real-world impact factors. For each factor, experiments were conducted on character-level recovery using the same sample sizes as specified above, i.e., 6,200 trajectories for content recovery and 100 sample pairs for biometric recovery.

\begin{figure}[!tbp]
  \centering
  \includegraphics[width=\linewidth]{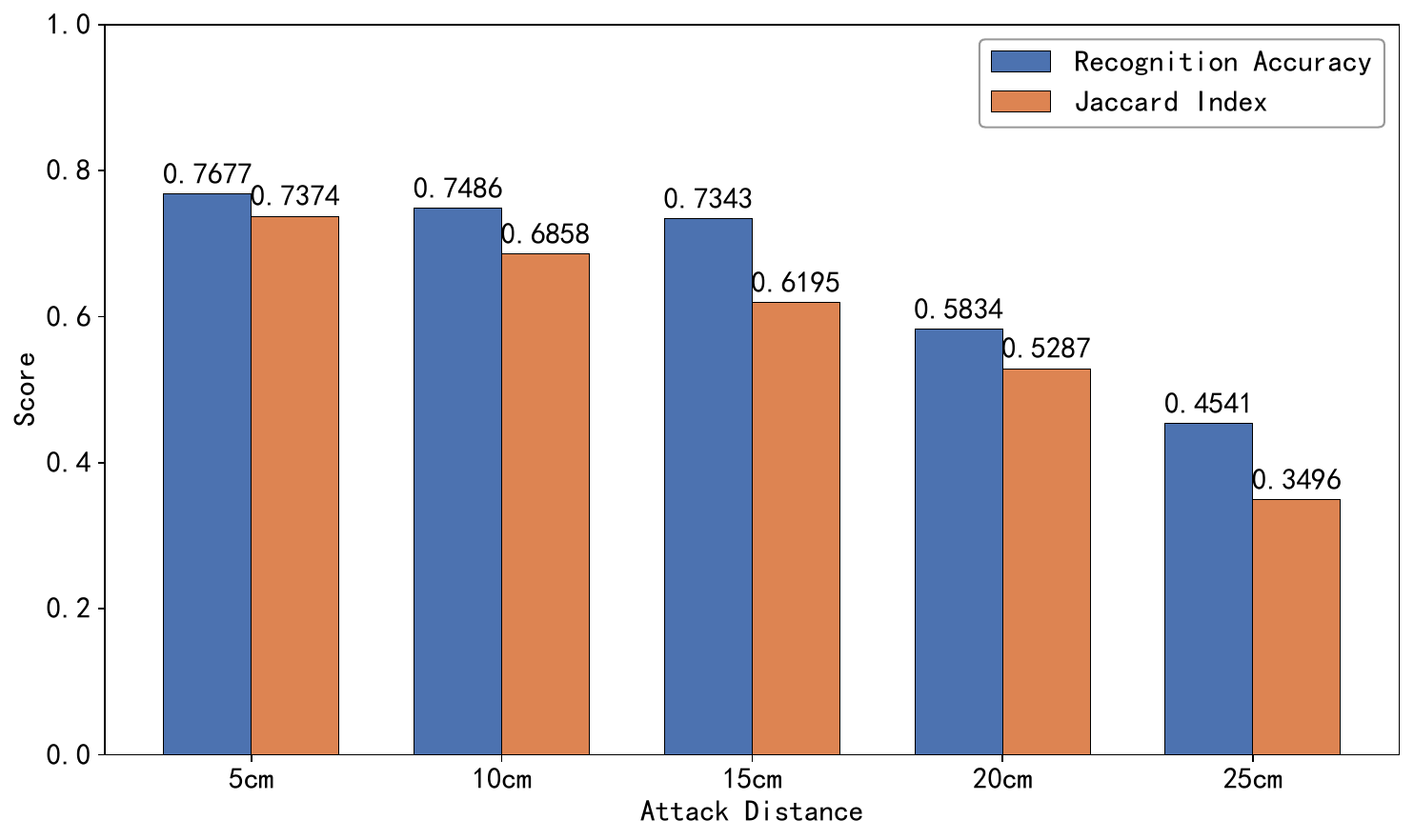}
  \caption{Impacts of probe-to-device distance.}\label{fig: distance}
  \vspace{0pt}
\end{figure}

\noindent\textbf{Evaluation on attack distance.} We first evaluated the impact of the probe-to-device distance, examining a range from 5 cm to 25 cm. As shown in Figure \ref{fig: distance}, attack performance gradually degrades as the distance increases. This degradation is a direct consequence of the physical properties of EM signals, which weaken with increasing distance, resulting in a reduced signal-to-noise ratio (SNR) at the probe. Despite this attenuation, the attack remains highly effective under realistic conditions. Even at 15 cm, \Name achieves a character recognition accuracy of 73\%, demonstrating that the attack is viable even when the probe is not in immediate proximity to the device.


\noindent\textbf{Evaluation on public and private Environments.}
We then assessed the attack's performance in two distinct real-world settings: a private meeting room and a public library. As described in Section \ref{sec: setup}, in the private scenario, the probe was concealed under a 5 cm thick wooden table. In the public scenario, it was placed under a bag at a distance of 15 cm from the device to simulate a more casual and electromagnetically noisy deployment. As shown in Table \ref{table: environment}, even in the noisy public environment, \Name still achieves a recognition accuracy of 72.49\% and a Jaccard index of 0.5946. These results demonstrate that the attack is not confined to controlled laboratory conditions and remains highly effective in realistic, noisy, and populated environments.
\begin{table}[t]
\centering
\caption{Impacts of public and private environments.}
\renewcommand{\arraystretch}{1}
\resizebox{\linewidth}{!}{
\begin{tabular}{c|c|c}
\Xhline{1.5pt}
\textbf{Environment}   & \textbf{Recognition Accuracy} & \textbf{Jaccard Index}   \\ \Xhline{1.5pt}
Private               &  0.7677            &                     0.7374                                                                                            \\ 
Public           & 0.7249 &                     0.5946 \\ 
\Xhline{1.5pt}

\end{tabular}}
\label{table: environment}
\end{table}

\noindent\textbf{Evaluation on different smartphone models}. Finally, we evaluated the generalizability of \Name across three additional smartphone manufacturers (i.e., Xiaomi 10 Pro, Samsung S10, and Huawei Mate 30 Pro) in the private scenario.  Variations in hardware design or display parameters may influence leakage EM signal characteristics. As shown in Table \ref{table: phonemodel}, the attack was effective against all tested devices, with single-character recognition accuracy exceeding 75\% and Jaccard index exceeding 0.68 for each model. This high level of consistency across different manufacturers strongly indicates that the underlying EM leakage is not an isolated design flaw but a systemic vulnerability inherently tied to the common hardware architectures used across the industry.

\begin{table}[t]
\centering
\caption{Evaluation results on different smartphones.}
\renewcommand{\arraystretch}{1}
\resizebox{\linewidth}{!}{
\begin{tabular}{c|c|c}
\Xhline{1.5pt}
\textbf{Smartphone}   & \textbf{Recognition Accuracy} & \textbf{Jaccard Index}   \\ \Xhline{1.5pt}
Xiaomi 10 Pro               &  0.7561            &                     0.7103                                                                                            \\ 
Samsung S10           & 0.7472 &                     0.6786 \\ 
Huawei Mate 30 Pro           & 0.8134 &                     0.7425 \\ 
\Xhline{1.5pt}

\end{tabular}}
\label{table: phonemodel}
\end{table}

\section{Conclusion}
In this paper, we reveal an unexplored EM side-channel vulnerability: the leakage of continuous, fine-grained handwriting trajectories from capacitive touchscreens. Based on this discovery, we propose \Name, a practical non-contact attack framework that captures these EM emanations and employs a transformer model to reconstruct the 2D path of a user’s writing. Our experimental evaluation confirms that \Name can recover handwriting trajectories from multiple COTS smartphones with high recognizability and geometric similarity, and remains robust under realistic environmental conditions. Our findings underscore the severity of this vulnerability and highlight an urgent need for the security community to develop effective countermeasures.
\clearpage
\bibliographystyle{ACM-Reference-Format}
\bibliography{sample-base}


\end{document}